\documentclass[11pt]{article} 
\usepackage{moriond,epsfig} 

\def\Journal#1#2#3#4{{#1} {\bf #2}, #3 (#4)} 
 
\def\NPA{{\em Nucl. Phys.} A} 
\def\PLB{{\em Phys. Lett.}  B} 
\def\PRL{\em Phys. Rev. Lett.} 
 
\def\PRC{{\em Phys. Rev.} C} 
 
\def\EPJC{{\em Eur. Phys. J} C}  
\def\JPG{{\em J. Phys.} G}  

\def\be{\begin{equation}} 
\def\ee{\end{equation}} 
\def\bea{\begin{eqnarray}} 
\def\eea{\end{eqnarray}} 
 
\begin{document} 
\vspace*{4cm} 
\title{NEW CERES RESULTS} 
 
\author{A.~MAR\'IN for the CERES COLLABORATION }
 
\address{Gesellschaft f\"ur Schwerionenforschung (GSI), Planckstr.~1, 64291 Darmstadt, Germany} 
 
\maketitle\abstracts{During 1999 the CERES experiment upgraded by the
new TPC was taking data from semicentral Pb+Au collisions at 40~AGeV. The 
analysis of the di-electron spectra shows an enhancement in 
the low mass region of at least the same magnitude as previously observed at full SPS 
energy. The results are compared to model calculations with and without
medium modifications of the vector mesons. The 
addition of the radial TPC gives more possibilities to study 
hadronic observables. The results obtained for the midrapidity $\Lambda$ yield and the 
$\overline{\Lambda}/\Lambda$ ratio are compared to the existing systematics.}

\section{Introduction} 
 
The CERES/NA45 experiment at CERN SPS is dedicated to the measurement of low-mass 
e$^+$e$^-$ pairs \cite{exp} in 
ultrarelativistic heavy-ion collisions. QCD calculations have predicted a 
transition from ordinary hadronic matter into a
plasma of deconfined quarks and gluons at high energy density. Dileptons, which have 
negligible final state interactions, represent a more
suitable probe than hadrons for the study of this new state of matter. CERES 
has observed an enhanced dilepton production in the invariant mass region 
$m_{e^+e^-}>$~0.2~GeV/c$^2$ in Pb+Au at 158 AGeV and in S+Au at 200~AGeV \cite{exp} 
compared to the contribution from known hadronic sources. The enhancement is 
not present on p-induced reactions \cite{pp}. Pion annihilation has been pointed out
as a possible mechanism for additional e$^+$e$^-$ production but the shape of the 
experimental spectra cannot be explained without introducing medium modifications of 
vector mesons, particularly of the $\rho$. The 
transition from normal hadronic matter to a state in which chiral
symmetry is restored is believed to be influenced by baryon density more strongly
than by temperature \cite{densi}. CERES was running during 1999 at 40~AGeV, thus, providing a second 
measurement at very different values of $\rho$ and T compared to 158~AGeV \cite{lowene1}. In 
addition hydrodynamical calculations showed that the softest point
is reached in  heavy ion collisions at around 30~AGeV \cite{lowene2}. Measuring at 
lower energies could be a strategy to search for the QCD phase transition.

We will focus in this article on the results from 1999 data taking concerning
dileptons \cite{hep,qm01} and $\Lambda$ \cite{sqm01} hyperons. A total of about 
8$\cdot$10$^6$ events with 30$\%$ centrality were recorded. Due to readout problems
in part of the detector this data set is limited in terms of statistics and momentum resolution. 

\section{Experimental Setup}
The CERES experiment is optimized to measure low mass electron pairs close to 
midrapidity (2.1$<\eta<$2.6) with full azimuthal coverage. In order to
improve the mass resolution to $\delta m/m<$ 2\% the CERES experiment
was upgraded in 1998 with the addition of a
cylindrical radial drift TPC and a new magnet system \cite{qm99}. A 
silicon telescope, composed of two silicon drift chambers (SDD) positioned at 10 cm 
and 13.8 cm behind a segmented Au target, provides a precise vertex reconstruction and 
angle measurement for charged particles. Two Ring Imaging CHerenkov (RICH) detectors 
operated at a high threshold ($\gamma_{th}$=32) are used for electron 
identification in a huge hadronic background. The new radial drift TPC
positioned downstream from the existing spectrometer has an active length of 2 m 
and an outer diameter of 2.6 m. It is operated inside a variable magnetic field with a 
maximal radial component of 0.5 T providing the measurement of up to 20
space points for each charged particle track. This 
is sufficient for the momentum determination and additional
particle ID via d$E$/d$x$. As particle ID and momentum 
measurement are separated in the upgraded apparatus, the two RICH
detectors can be used in a combined mode, resulting in an improved 
electron efficiency (from 0.70 in 1995/96 to 0.94 in 1999) and improved rejection power. Moreover, 
the addition of the radial TPC gives wider possibilities to study hadronic observables. 

\section{Electron analysis}

Electrons are identified by the ring pattern detected in the RICH
detectors. An electron track is constructed by matching RICH rings to the track segments 
identified in the SDD and TPC. Invariant mass is calculated
by pairing oppositely charged electron tracks. Although the RICH
detectors reject 95$\%$ of all hadrons and the total detector material
is kept as low as $X/X_0$=1$\%$ the main problem of an electron analysis is the combinatorial
background due to photons from conversions and $\pi^0$ Dalitz decays
and to remaining hadrons. 

\begin{figure}[hb]
\begin{minipage}[t]{60mm}
\includegraphics*[width=7.cm]{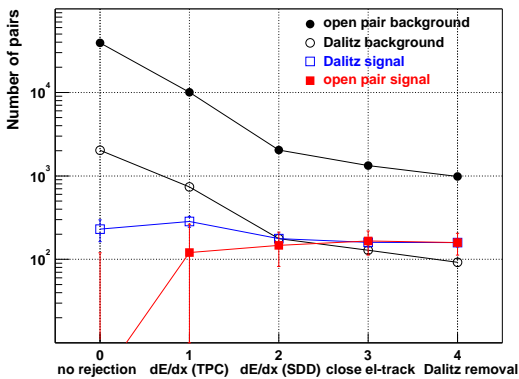}
\end{minipage}
\begin{flushright}
\begin{minipage}[t]{55mm}
\vspace{-5cm}
\caption{Evolution of the total number of pairs after various
rejection steps, shown for the open pair signal 
(m$_{ee}>$~0.2~GeV/c$^2$), open pair background (m$_{ee}>$0.2~GeV/c$^2$), 
Dalitz signal (m$_{ee}<$~0.1~GeV/c$^2$) and Dalitz background 
(m$_{ee}<$~0.1~GeV/c$^2$). }
\label{fig:electroncuts}
\end{minipage}
\end{flushright}
\end{figure}

\begin{figure}[ht]
\includegraphics*[width=6.5cm]{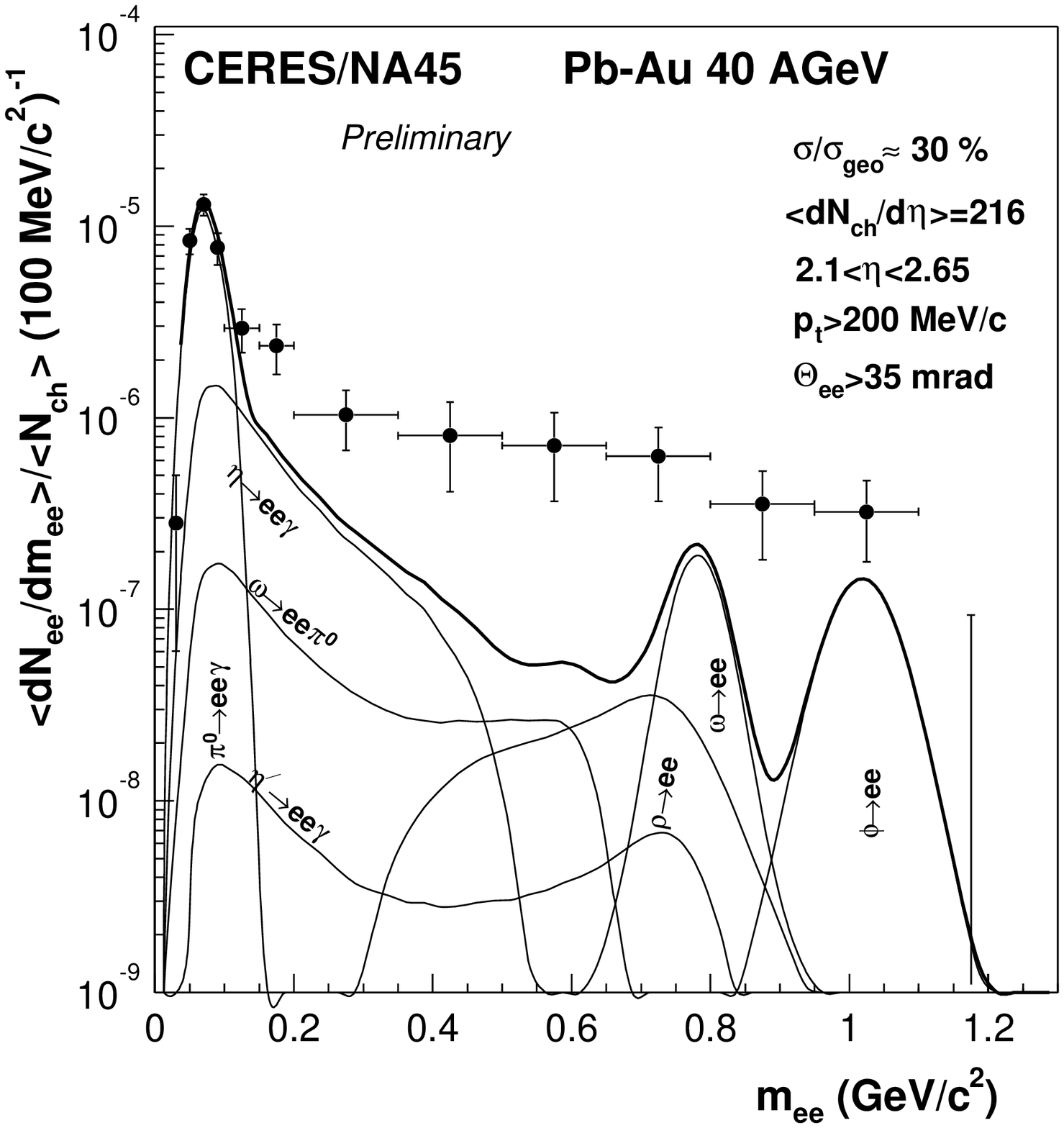}
\includegraphics*[width=6.5cm]{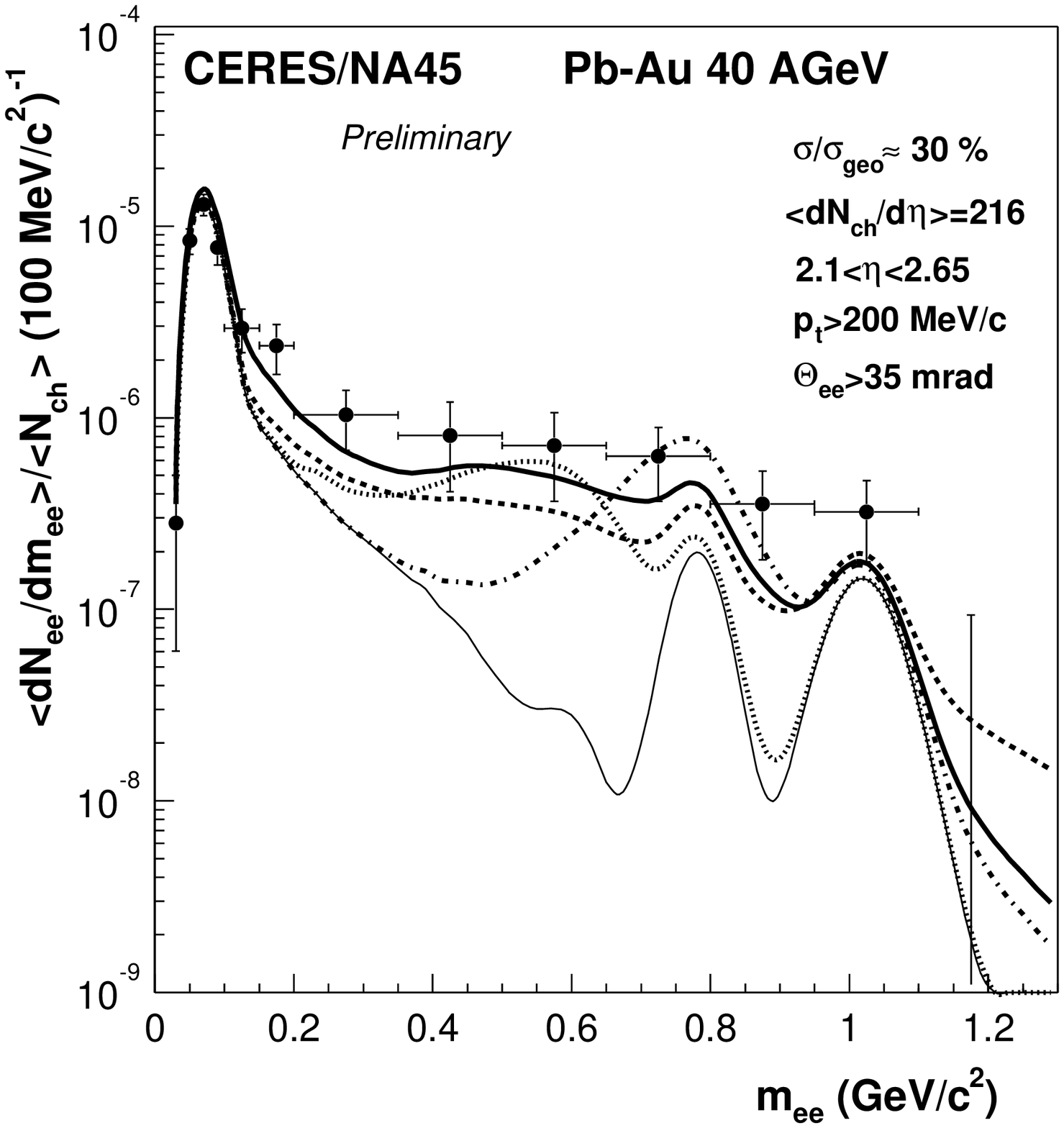}
\caption{Invariant mass spectrum  of $e^+e^-$ pairs at 40~AGeV. Left:
Comparison to the expectations from known hadronic sources in heavy ion 
collisions (``decay cocktail''). Right: Comparison of the data to model
calculations, based on $\pi^+\pi^-$ annihilation with an unmodified
$\rho$ (dashed-dotted), a dropping $\rho$ mass (dotted) and broadening of the
$\rho$ spectral function (thick solid). Also shown is the hadronic cocktail without $\rho$ (solid). 
The lowest order pQCD rate calculation (dashed). }
\label{fig:electroninvmass}
\end{figure}

A few experimental cuts applied in the analysis provide a rejection factor larger than 100 with 
efficiency losses for open pairs less than a factor 2 (extracted from Dalitz signal losses), as illustrated in 
Fig.~\ref{fig:electroncuts}. Only electron tracks with transverse momenta $p_t>0.2$ GeV/c and with an
opening angle $\theta_{e^+e^-}>$ 35 mrad are taken (Fig.~\ref{fig:electroncuts}, cut 0). Additional 
electron identification and pion rejection is achieved by a
cut in d$E$/d$x$ in the TPC as a function of the momentum (Fig.~\ref{fig:electroncuts}, cut 1). Conversions 
and Dalitz pairs with too small opening angle to create two separate RICH rings produce a
double energy loss within small angular region in the SDD and can be removed by 
applying an upper cut in SDD d$E$/d$x$ (Fig.~\ref{fig:electroncuts}, cut 2). Due 
to the incomplete azimuthal readout of the TPC in 99 data set electron tracks that have a 
second SDD-RICH candidate within 70 mrad are also rejected (Fig.~\ref{fig:electroncuts}, cut 3). 
Positively identified Dalitz pairs are excluded for
further combinatorics (Fig.~\ref{fig:electroncuts}, cut 4). The final dilepton spectrum
(Fig.~\ref{fig:electroninvmass}) is obtained by 
subtracting the like-sign pairs from the unlike-sign pairs 
as N$_{e^+e^-}$-2(N$_{e^+e^+}\cdot$ N$_{e^-e^-}$)$^{1/2}$.

A total of 185$\pm$48 open pairs for $m_{e^+e^-}>$~0.2~GeV/c$^2$ with a signal to background
ratio of S/B~=~1/6 is obtained from the analysis of 40~AGeV data. Comparing 
to the hadronic decay cocktail an enhancement factor of 5.1$\pm$1.3 (stat.) is obtained. The 
pair $p_t$ distributions for two regions of mass $m_{e^+e^-}<$~0.2~GeV/c$^2$
and $m_{e^+e^-}>$~0.2~GeV/c$^2$ shows that the enhancement is localized at low $p_t$
in agreement with the observation at 158 AGeV. From the comparison to 
theory \cite{theory} (Fig.~\ref{fig:electroninvmass}) one concludes 
that strong medium modifications of the intermediate $\rho$ are needed
to explain the data but the accuracy of data does not allow to distinguish 
between different scenarios, $\rho$ scaling reducing the mass or broadening 
of the $\rho$ spectral function. The enhancement is consistent or even larger than the 
one observed at 158 AGeV showing that baryons play the dominant role in the medium modifications 
of the $\rho$.

\section{$\Lambda$ production}

The addition of the radial TPC gives more possibilities to study 
hadronic observables in the CERES experiment. For example, it allows a
systematic investigation of the $\Lambda$ hyperon at midrapidity. The 
$\Lambda$ hyperon is reconstructed by the invariant mass of
positive and negative particles in one event \cite{sqm01}. Only the information
of the TPC and the SDD detector are used in
this analysis. Positive particles are taken in the transverse momentum range
0.5~$<p_t<$~2.0 GeV/c and negatives in the range 0.25~$<p_t<$~0.6~GeV/c.
The final $\Lambda$ acceptance is 0.9~$<p_t<$~2.5~GeV/c and 2.0~$<y<$~2.4.
A cut on the Armenteros Podolanski plot of 10-120~MeV/c in $p_t^+$ 
and $\alpha<$0.65 is applied. Three analysis methods were used: first, only TPC information is
required. Second, TPC tracks should not have a matched track in the SDD, in that way
only $\Lambda$'s decaying at least 13 cm after the vertex are detected. Third, after reconstructing 
the decay point an additional cut of 50 cm in the decay point is applied.
\begin{figure}[ht]
\includegraphics*[height=6.6cm]{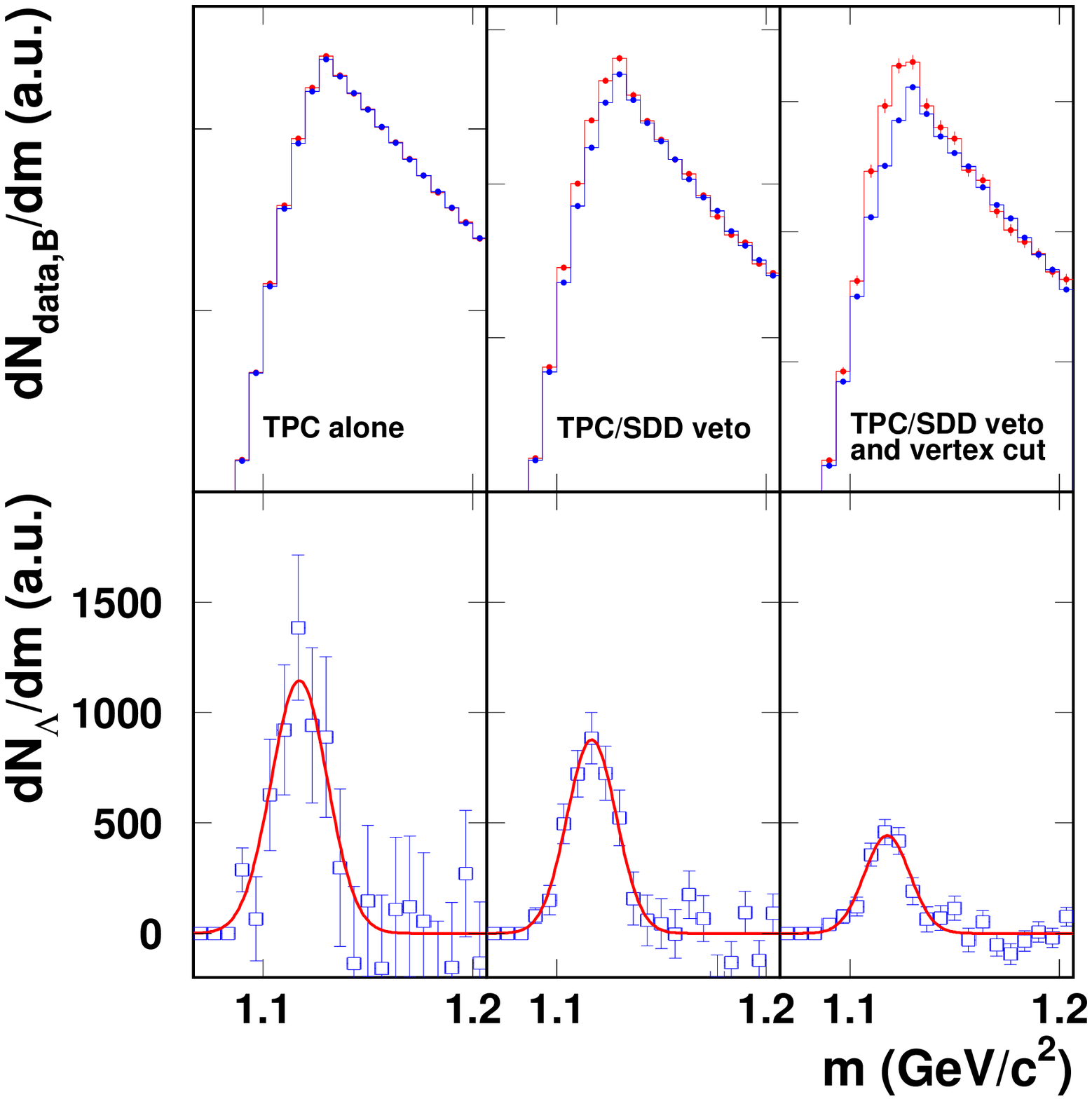}
\includegraphics*[height=6.7cm]{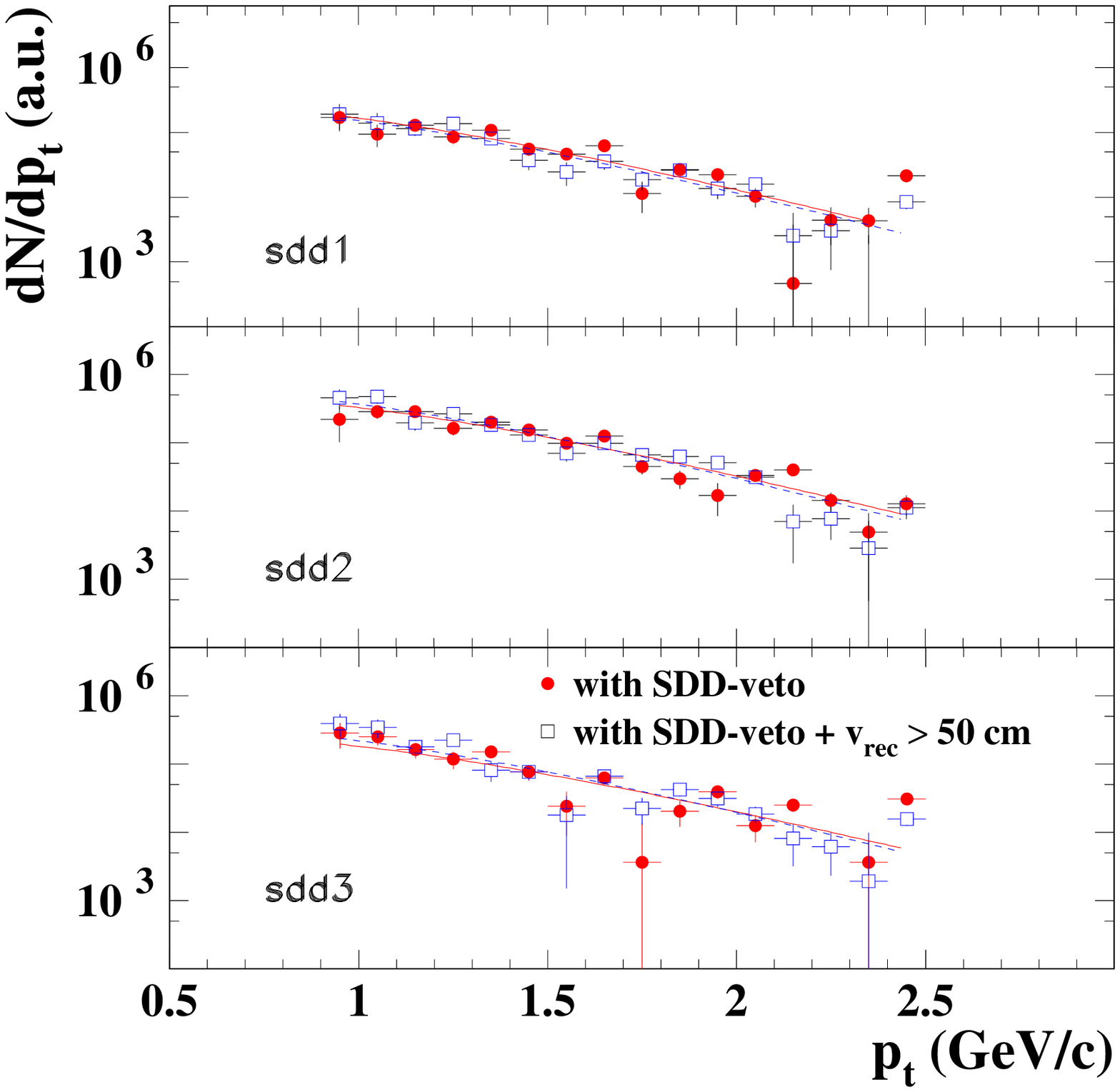}
\vspace{-0.2cm}
\caption{Left: Invariant mass of positive and negative particles (dashed line) and 
combinatorial background (solid line) for the three analysis
methods. In the bottom is shown the signal after background 
subtraction. Right: Transverse momentum spectra for $\Lambda$ shown separately for two different
analysis methods and for the three centrality classes in the rapidity region 2$<y_\Lambda<$2.4. All 
spectra are acceptance corrected but not normalized to the number of events.}
\label{fig:lambdainvmass}
\vspace{-0.6cm}
\end{figure}
The invariant mass spectra for the three different methods assuming positives to be protons and
negatives to be pions is shown in
Fig.~\ref{fig:lambdainvmass} (left). Although one can notice a loss of
signal, a large improvement in signal to background is obtained as
successive cuts are introduced. After subtraction 
of the combinatorial background the $\Lambda$ signal is obtained at a mass 
of $m_\Lambda$=1.117$\pm$0.001 GeV/c$^2$ close to the PDG value. The width 
$\sigma_\Lambda$ =11.2$\pm$0.4 MeV/c$^2$ is larger than
that expected from the design momentum resolution due to the incomplete 
calibration process. As consistency 
check the number of $\Lambda$'s as a function of the
decay point is fitted to an exponential function giving a c$\tau_0$ of 
(8.1$\pm$0.5) cm, in agreement with the tabulated value of 7.89 cm. Transverse 
momentum distributions are depicted in Fig.~\ref{fig:lambdainvmass} (right). An 
inverse slope parameter $T$ of 273$\pm$20 MeV is obtained from a fit to an 
exponential function $dN/p_tdp_t \propto exp(-m_t/T)$, slightly increasing with centrality.

\begin{figure}[h]
\begin{minipage}[t]{60mm}
\vspace{-0.5cm}
\includegraphics*[height=5.5cm]{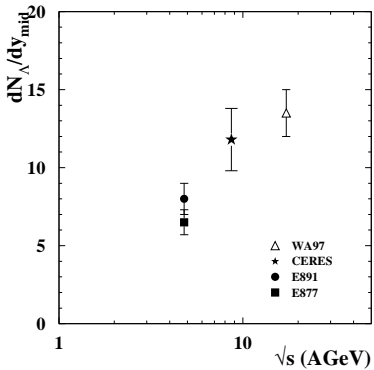}
\vspace{-0.6cm}
\caption{$dN_{\Lambda}/dy_{mid}$ as a funtion of beam energy $\sqrt{s}$. } 
\label{fig:lambdayield}
\end{minipage}
\end{figure}
\begin{flushright}
\begin{minipage}[t]{85mm}
\vspace{-6.4cm}
A midrapidity density 
of $dN_{\Lambda}/dy_{mid} = 11.8\pm 2.0$ is obtained by extrapolating the transverse 
momentum spectra to $p_t$=0 using the fitted function. In addition a ratio 
of $\overline{\Lambda}/\Lambda$= 0.024$\pm$0.010 has been obtained, in agreement 
with the result reported by NA57 \cite{sps1}.

When comparing the midrapidity  $\Lambda$ yield (Fig.~\ref{fig:lambdayield}) 
and $\overline{\Lambda}/\Lambda$ ratio \cite{sqm01}
to the results at AGS \cite{ags1,ags2} and full SPS \cite{sps2} one observes a monotonic rising behaviour with increasing
beam energy.
\end{minipage}
\end{flushright}


\vspace{-1.5cm}

\section*{References} 
\vspace{-0.3cm}
\begin{thebibliography}{99} 

\bibitem{exp} G.~Agakichiev {\it et al.} CERES Collaboration \Journal{\PRL}
   {75} {1272}{1995};  \Journal{\PLB} {422} {405} {1998}; 1999 \Journal{\NPA} {661}{23c}{1999}.

\bibitem{pp} G.~Agakichiev {\it et al.} \Journal{\EPJC}{4}{231}{1998}. 

\bibitem{densi} S.~Klimt, M.~Lutz and W.~Weise \Journal{\PLB}{249}{386}{1990}.

\bibitem{lowene1} G.E.~Brown and M.~Rho \Journal{\PRL}{66}{2720}{1991}; 
T.~Hatsuda and S.~Lee \Journal{\PRC} {46} {R34}{1992};
M.~Herrmann, B.~Friman, W.~N\"orenberg \Journal{\NPA}{560}{411}{1993};
K.Haglin, \Journal{\NPA}{584}{719}{1995}.

\bibitem{lowene2} C.M.~Hung and E.V.~Shuryak \Journal{\PRL}{75(22)}{4003}{1995}.

\bibitem{hep} S.~Damjanovi\'c and K.~Filimonov. Proc. of Int. Europhys. Conf. in HEP, Budapest, 2001.

\bibitem{qm01} H. Appelsh\"auser for the CERES Collaboration \Journal{\NPA}{698}{253c}{2002}.

\bibitem{sqm01} W. Schmitz for the CERES Collaboration SQM01 (to be published) \Journal{\JPG} {}{}{}.

\bibitem{qm99} A. Mar\'{\i}n for the CERES Collaboration \Journal{\NPA} {661}{673c}{1999}.

\bibitem{theory} R. Rapp. Private communication.
\bibitem{sps1} N.~Carrer for the NA57 Collaboration \Journal{\NPA}{698}{118c}{2002}.
\bibitem{ags1} S.~Ahmad {\it et al.} E891 Collaboration \Journal{\PLB}{382}{35}{1996}.
\bibitem{ags2} K.~Filimonov for the E877 Collaboration  \Journal{\NPA}{661}{198c}{1999}.
\bibitem{sps2} F.~Antinori {\it et al.} WA97 Collaboration \Journal{\EPJC}{11}{79}{1999}.

\vspace{-2.5cm} 
\end{thebibliography}
\end{document}